\documentclass{aa}
\usepackage{psfig}
\begin{document}
\title{New measurement of the $^6$Li/$^7$Li isotopic ratio in the extra-solar 
planet host star HD82943 and line blending in the Li 6708 \AA\ region.}
\author{G.~Israelian\inst{1}
\and 
 N.~Santos\inst{2}
\and
M.~Mayor\inst{2}
\and
R.~Rebolo\inst{1}
}
\offprints{G.Israelian, \email{gil@iac.es}} 
\institute{Insituto de Astrofisica de Canarias, E-38205 La Laguna, 
Tenerife, Spain
\and 
Observatoire de Gen\`eve, 51 ch.  des 
Maillettes, CH--1290 Sauverny, Switzerland
}
\date{Received; accepted}

\titlerunning{A new measurement of the $^6$Li/$^7$Li  ratio in HD\,82943}

\abstract{
The presence of possible blends in the spectral region of the Li resonance line at 6708 \AA\ 
in solar-type metal-rich stars is investigated using high resolution and high signal-to-noise 
spectroscopic observations. Our analysis does not confirm the identification of a 
weak absorption feature at 6708.025 \AA\ with the low excitation \ion{Ti}{i} line 
proposed by Reddy et al.~(2002). Our spectrum synthesis suggests that the 
unidentified 
absorption is most probably produced by a high excitation \ion{Si}{i} line 
originally proposed by M\"{u}ller et al.~(1975). Reanalysis of the $^6$Li/$^7$Li 
isotopic ratio in HD\,82943 was performed by taking the \ion{Si}{i} line into account 
and using new VLT/UVES spectra of HD\,82943 with a signal-to-noise ratio close to 
1000. 
We confirm the presence of $^6$Li in the star's atmosphere while the updated 
value for the isotopic ratio is $f(^6$Li) = $0.05\pm0.02$.

}

\maketitle

\section{Introduction}

An interesting opportunity for testing the planet(s) ingestion
scenario is offered by a $^6$Li test (Israelian et al.~\cite{isr01}).
This approach is based on looking for an element that should not appear 
in the atmosphere of a normal star, but would be present in a star that 
has swallowed a planet. We have proposed recently that a light 
isotope, $^6$Li, is an excellent tracer of any planetary matter
accretion. The detection of $^6$Li in HD\,82943 by our group 
(Israelian et al.~\cite{isr01}) was considered as convincing observational evidence 
that stars may accrete planetary material, or even entire planets, during their 
main sequence (MS) lifetime. Other explanations (such as stellar flares 
or surface sports) of this phenomenon were ruled out (Israelian et al.~\cite{isr01}). 
It has been suggested (Sandquist et al.~\cite{san02}) that $^6$Li  can  be used 
to distinguish between different giant planet formation theories.

Nuclear reactions destroy the lithium isotopes ($^6$Li and $^7$Li) in
stellar interiors at temperatures $2\times10^6$ ($^6$Li) and $2.5\times10^6$ K
($^7$Li). 
Convection cleans the upper atmosphere of Li nuclei by transporting them to deeper,
hotter layers where they are rapidly destroyed. Young low mass stars are entirely 
convective and most of the primordial Li nuclei are burned in their interiors 
in a mere few million years. However, many solar-type  
stars preserve a large fraction of their initial atmospheric $^7$Li nuclei.
According to standard models (Forestini \cite{for94}; 
Montalb\'an \& Rebolo \cite{mon02}), at a given metallicity there is a mass
range, where $^6$Li but not $^7$Li is being destroyed. These models predict
that no $^6$Li can survive pre-MS mixing in metal-rich solar-type stars.
Observations seem to support this scenario since no clear detections of 
$^6$Li have been reported in solar-type solar-metallicity stars 
(Andersen, Gustaffson \& Lambert \cite{and84}; Rebolo et al.~\cite{reb86}). 
We note, however, that the mass and the depth of the convection zone 
also depend on the metal content of the star, and for this reason several 
old metal-poor stars have preserved a fraction of their initial $^6$Li nuclei 
(Cayrel et al.~\cite{cay99}, Hobbs, Thorburn \& Rebull \cite{hobbs99}). 
This suggests that any detected $^6$Li in a metal-rich solar-type star would most
probably signal an external source for this fragile isotope. 

Observations of $^6$Li are extremely difficult for several reasons. First
of all, it is a weak component of a blend of the much stronger doublet 
of $^7$Li with an isotopic separation 0.16 \AA. In the case of metal-poor 
halo stars with [Fe/H]$< -1$, blending of the Li line with other weak
absorptions is not expected, and also the placement of
continuum does not pose serious problems. $^6$Li has been unsuccessfully 
sought in many metal-poor stars but unambiguously detected  in only few 
(Hobbs et al.~\cite{hobbs99}; Cayrel et al.~\cite{cay99}; Nissen et al.~\cite{Nis99}). 
The methods used in the analysis of $^6$Li have 
been widely discussed in the literature (e.g. Nissen et al.~\cite{Nis99}; 
Cayrel et al.~\cite{cay99}; Hobbs et al.~\cite{hobbs99}).
In metal-rich stars the identification of any possible weak blends in the
region of the Li absorption becomes crucial (King et al.~\cite{king97}). 
Recently Reddy et al.~(\cite{red02}) updated the 
line list of Lambert, Smith \& Heath (\cite{lam93}) based on the analysis 
of the solar spectrum of Kurucz et al.~(\cite{kur84}) and Hinkle et 
al.~(\cite{Hin00}). They claimed that a previously noticed weak absorption
in the solar spectrum at 6708.025 \AA\  (M\"{u}ller, Peytremann 
\& De La Reza \cite{mull75}) belongs to \ion{Ti}{i}. 
With this assumption their analysis of the Li
feature in HD\,82943 did not confirm the presence of $^6$Li.
 
With the goal of establishing the nature of the absorption at 
6708.025 \AA\  and better understanding the line list near Li, 
we have obtained high resolution, high S/N spectra for 
several metal-rich stars with no detectable Li line. 
Our targets cover a wide range of effective temperature, allowing 
 a more reliable identification of the lines in the spectral region
of the Li feature. We have also obtained new spectra of HD\,82943. 
Here we present new analysis of the $^6$Li/$^7$Li ratio in HD\,82943 and 
critically discuss the line list of the Li region. Our observations do not confirm 
the \ion{Ti}{i} line of Reddy et al.~(\cite{red02}) and support our previous
claim for detection of $^6$Li.

\section{Observations}

We used the 8.2 m VLT Kueyen (ESO, La Silla), the UVES spectrograph and an 
EEV detector to obtain spectra of HD\,82943 during the night of 2001 
March 5. The 0.3\arcsec\ slit yielded a resolving power of $R \sim$ 105\,000 
as measured from the FWHM of Th--Ar lamp lines. 

The 4.2 m William Herschel Telescope on La Palma (Canary Islands, 
Spain) and the UES spectrograph were used to observe several 
solar-type stars with large Li depletions in their atmospheres. 
Our first targets were HD\,171888 (F8 V), HD\,4747 (G8 V), HD\,217580 (K4 V),
HD\,217107 (G3 V) and HD\,210277 (G4 V). These stars were observed 
on 2000 July 8. The CCD SITE1 (2148 $\times$ 2148 pixel), 
grating E31, slit width 1\arcsec\  (providing a resolving power of 55\,000) 
and the central wavelength of 5140 \AA\ were used during this run. Our 
spectra had an S/N ratio  in the range 300--350, which was sufficient
to guarantee the detection of any spectral lines with a minimum 
equivalent width  of $\sim$1.5 m\AA. These data were used by us 
(Israelian et al.~2001)
in order to check the presence of blends near 6708 \AA\ with the EW 
larger than 1.5 m\AA. More targets (Table 1) have been observed 
in order to reduce this limit to below 1 m\AA. The same configuration was 
used in  2001 October to observe HD\,16141, HD\,22049, HD\,75732A and HD\,217107. 
In the next observing run (2002 June 20), we used a $4096\times 2048$-pixel EEV CCD   
with a pixel size  of 15 $\mu$m, the E31 grating, a slit width  of 1\arcsec\  and a 
central wavelength of 7215 \AA. Ten exposures of HD\,82943 were taken on 
2001 February 5 with a mosaic of two EEVs, the  E79 grating and a slit width of
0.7\arcsec\  in order to obtain a high quality of spectrum of this star.

The SARG high resolution echelle spectrograph at the 3.5 m Telescopio Nazionale 
Galileo (La Palma) was used during two runs in  2001 August and  2002 May.
The spectra were obtained with the yellow grism and span the wavelength 
range 4600--7820 \AA\ at a resolving power of $\sim$57\,000. The CCD was a 
mosaic of two  $4096\times 2048$-pixel EEVs with a pixel size of
15 $\mu$m.

All our   VLT/UVES, WHT/UES and TNG/SARG spectra were reduced using 
standard IRAF\footnote{IRAF is distributed by National Optical Astronomy 
Observatories, operated by the Association of Universities for Research 
in Astronomy, Inc., under contract with the National Science Foundation, 
USA.} routines. Normalized flats created for each observing night were 
used to correct the pixel-to-pixel variations and a Th--Ar lamp was used 
to find a dispersion solution.

Some observations were made with the FEROS spectrograph 
at the ESO 1.52 m telescope in La Silla. The mosaic of two $4096\times 2048$-pixel
EEV CCDs  were used to observe HD\,1461 and HD13\,445 on 2000 November 10 
and HD\,192263 on 2001 October 31.  The spectra were flatfielded, 
calibrated with a Th--Ar lamp and reduced using MIDAS routines.

\begin{table*}
\caption[]{Observing log and stellar parameters adopted from the references listed in 
the last column: 
(1) Santos et al.~(2002),
(2) Santos et al.~(\cite{San01}), 
(3) Santos et al.~(2003), 
(4) Mallik (\cite{mal99}),
(5) this article, 
(6) Boesgaard et al.~(\cite{boes01}). The S/N is calculated near 6706.5 \AA. }
\begin{tabular}{lccccccccccc}
\hline
\noalign{\smallskip}
Target  & Configuration & $V$ & Exp.~time & $\lambda$/$\Delta \lambda$ & S/N &Date 
& $T_\mathrm{eff}$ & $\log{g}$   &  $\xi_{t}$  & [Fe/H] & Ref. \\
        &               &     & [s]    &     &     &     & [K]  &  [cm\,s$^{-2}$] & 
	[km\,s$^{-1}$] &  &  \\
	\hline \\
HD\,1461   & 1.52m/FEROS & 6.46 & 1200 & 48,000 & 400 & 2000 Nov 10 & 5785 & 4.47 & 1.23 & 0.18  & 1\\
HD\,13445  & 1.52m/FEROS & 6.10 & 900  & 48,000 & 350 & 2000 Nov 10 & 5190 & 4.71 & 0.78 & $-$0.20 & 2\\
HD\,16141  & WHT/UES     & 6.78 & 8100 & 55,000 & 600 & 2001 Oct 03 & 5805 & 4.28 & 1.37 & 0.15  & 1\\
HD\,22049  & WHT/UES     & 3.73 & 700  & 55,000 & 800 & 2001 Oct 03 & 5135 & 4.70 & 1.14 & $-$0.07 & 2\\
HD\,75732A & WHT/UES     & 5.95 & 1200 & 55,000 & 400 & 2001 Oct 01 & 5307 & 4.58 & 1.06 & $+$0.35 & 3\\
HD\,82943  & VLT/UVES    & 6.54 & 700  & 105,000 & 900 & 2001 Mar 05 & 6025 & 4.53 & 1.10 & $+$0.30 & 3\\
HD\,82943  & WHT/UES     & 6.54 & 8400 & 70,000 & 1100 & 2001 Feb 05  & -  & -  & -  & -  & -  \\
HD\,130322 & TNG/SARG    & 8.0  & 3600 & 57,000 & 300 & 2002 May 25 & 5430 & 4.62 & 0.95 & $+$0.06 & 2 \\
HD\,134987 & TNG/SARG    & 6.45 & 1800 & 57,000 & 300 & 2002 May 25 & 5780 & 4.45 & 1.06 & $+$0.32 & 3 \\
HD\,134987 & WHT/UES     & 6.45 & 2000 & 55,000 & 300 & 2002 Jun 20 & -    & -    &   -  & -       & - \\
HD\,137759 & TNG/SARG    & 3.31 & 150  & 57,000 & 550 & 2002 May 26 & 4553 & 2.74 & 1.50 & $+$0.03 & 4 \\
HD\,145675 & TNG/SARG    & 6.67 & 19\,800 & 57,000 & 900 & 2002 May 27 & 5255 & 4.40& 0.68 & $+$0.51 &3 \\
HD\,177830 & WHT/UES     & 7.17 &  800 & 55,000 & 300 & 2000 Jul 09 & 4840 & 3.60 & 1.18 & $+$0.32 & 3 \\
HD\,192263 & 1.52m/FEROS & 8.1  & 1800 & 48,000 & 300 & 2001 Oct 31 & 4995 & 4.76 & 0.90 & $+$0.04 & 3\\
HD\,198084 & WHT/UES     & 4.51 & 1500 & 55,000 & 800 & 2002 Jun 20 & 6171 & 4.13 & 1.90 & $+$0.15 & 6 \\
HD\,200790 & WHT/UES     & 5.97 & 2400 & 55,000 & 600 & 2002 Jun 20 & 6240 & 4.26 & 1.70 & $+$0.02 & 5\\
HD\,210277 & TNG/SARG    & 6.63 & 3000 & 57.000 & 400 & 2001 Aug 02 & 5560 & 4.46 & 1.03 & $+$0.21 & 3 \\
HD\,210277 & WHT/UES     & 6.63 & 1400 & 55,000 & 400 & 2002 Jun 20 & -    & -    & -    & - & - \\
HD\,217107 & WHT/UES     & 6.18 & 4800 & 55,000 & 800 & 2001 Oct 03 & 5655 & 4.42 & 1.11 & $+$0.38 & 2 \\
\noalign{\smallskip}
\hline
\end{tabular}
\label{tab1}
\end{table*}

\section{Spectrum synthesis and stellar parameters}

In this article we used models of atmospheres provided by Kurucz (1992) and
the spectrum synthesis code MOOG (Sneden 1973). We have compared abundances of
various elements computed with our atmospheric models and with those used by Reddy et 
al.~(1999) (i.e. Kurucz 1995). We found that differences induced in the 
abundances by the two types of models are less than 0.01 dex and  can therefore be neglected. 
Solar abundances of chemical elements were taken from Anders \& Grevesse 
(\cite{And89}). We have also made an extensive use of the VALD database (Kupka et al.~1999). 
The parameters of some of the planet host stars listed in Table 1 were slightly updated with 
respect to the values given in Santos et al.~(2001, 2002). Three stars in our sample 
(HD\,1461,  HD\,198084 and HD\,200790) are not known to harbour planets and  
were observed because of the absence of Li. We have attempted to
estimate the parameters of a cool subgiant, HD\,137759, from the set of
Fe lines used in our previous analysis (Santos et al.~\cite{San01}).
However, given the low temperature of this object, our final values
had large error bars and we have decided to use the parameters based
on colours and accurate parallax measurement from Hipparcos (Mallik 1999).
Two Li-poor dwarfs, HD\,198084 and HD\,200790, were taken from
Boesgaard et al.~(\cite{boes01}). The parameters of HD\,200790 were estimated
in the same way as for the other stars in our sample. HD\,198084 is a metal-rich
F8IV-V (Boesgaard et al.~\cite{boes01}, Malagnini et al.~\cite{mal00}) 
spectroscopic binary (Griffin \cite{grif90}). 
The iron lines in our spectrum were strongly blended owing to the presence 
of the second star and we did not make a new estimate of the stellar parameters.

\section{The line list in the Li region}

Accurate measurements of the wavelengths and  oscillator strengths of the Li lines
are available in the literature (Sansonetti et al.~1995).
Unidentified blends in the Li region of the Sun and other cool stars have been
discussed by several authors (M\"{u}ller, Peytremann \& De La Reza \cite{mull75}; 
Lambert, Smith \& Heath \cite{lam93}; Andersen et al.~\cite{and84}; 
King et al.~\cite{king97}).
M\"{u}ller et al.~(\cite{mull75}) were the first to notice the weak feature at 
6708.025 
\AA\ in the solar spectrum and attributed it to the \ion{Si}{i} line with 
$\log gf=-2.7$, $\chi$ = 6 eV. The line list of Lambert et al.~(\cite{lam93}) has 
been extensively discussed and modified by King et al.~(\cite{king97}). 
Following M\"{u}ller et al.~(1975), 
these authors used a fictitious high excitation \ion{Si}{i} line with 
$\chi_{l}$ =6 eV in order to account for the weak absorption in the 
spectrum synthesis. 

The line was also considered by Nissen et al.~(1999) 
in their analysis of metal-poor disc stars. According to Nissen et al.~(1999), 
the equivalent width (EW) of the  6708.025 \AA\  absorption is 0.6 m\AA\ in the solar 
spectrum. Reddy et al.'s (\cite{red02}) identification of this absorption with the 
Ti line 
($\log gf=-2.252$, $\chi$ = 1.88) implies a slightly different value EW = 0.75 m\AA. 
While this feature is too weak to affect the determination of the Li isotopic 
ratio in metal-poor halo stars (where it certainly disappears), it may 
cause problems in solar metallicity or more metal-rich stars.
Apart from this unidentified feature, there are other weak lines of 
\ion{Cr}{i} 6707.64 \AA, \ion{Ce}{i} 6707.74 \AA\ and CN 6707.816 \AA\  
with uncertain $gf$ values. It is interesting that the best synthetic spectra fits 
to the Li region of different stars can be achieved only by altering the $gf$ values 
of these weak lines by different amounts. This led King et al.~(\cite{king97}) 
to propose 
that various uncertainties in the analysis of the solar spectrum 
allow Li isotopic ratios as high as 0.1. 

Most of the lines in the list of Reddy et al.~(\cite{red02}) come from the original 
compilation of Lambert et al.~(\cite{lam93}). However, the \ion{Ti}{i} 6707.752 \AA\  
line was not listed in any of the references cited by Reddy et al.~(\cite{red02}). 
On the other hand, the \ion{V}{i} 6707.563 \AA\ line in the list of Reddy et 
al.~(\cite{red02})
appears under 6707.518 \AA\ in the VALD-2 database (Kupka et al.~1999). 
The $gf$ values of many lines have been modified by Reddy et al.~(\cite{red02}) 
in order to fit the solar spectra of Kurucz et al.~(1984), Hinkle et al.~(2000) and 
16 Cyg B. In most cases there is a clear disagreement with the list 
of King et al.~(\cite{king97}), who employed 
the same spectrum synthesis code, model atmospheres and, as a matter of fact, 
the same targets (i.e.~the Sun and 16 Cyg B). Despite this, King et al.~(\cite{king97}) 
found for CN 6707.816 \AA\  $\log gf=-1.74$ while Reddy et al. (\cite{red02}) 
arrived at $\log gf=-2.317$. Futhermore, King et al.~(\cite{king97}) have 
considerably increased the $\log gf$ value of a \ion{Ce}{ii} line at 6707.74 \AA\  and 
obtained $\log gf=-1.86$ while Reddy et al.~(\cite{red02}) decreased it to
$-$0.8 dex after 
Lambert et al.~(1993) and obtained a value $\log gf=-3.81$. The \ion{V}{i} 
6708.07 \AA\  line with $\log gf=-2.33$ from Lambert et al.~(\cite{lam93}) appears 
under \ion{V}{i} 6708.094 \AA\ with $\log gf=-3.113$ in the list of 
Reddy et al.~(\cite{red02}). 
The latter took it from the list available in the web site of R. Kurucz. 
The $gf$ values of the \ion{Cr}{i} 6707.64 \AA\  line in King et al.~(1997) and 
Reddy et al.~(2002) differ by 0.14 dex. The source of these discrepances was not 
investigated by Reddy et al.~(\cite{red02}). 

Our own synthesis of the solar spectrum suggests $gf$ values  close to those proposed 
by King et al.~(\cite{king97}). For example, when doing a synthesis of $^6$Li in 
HD\,82943 (Israelian et al.~2001) we set $\log gf=-1.81$ for the CN 6707.816 \AA\ line,
which is similar to $\log gf=-1.74$, as found by King et al.~(1997). In addition to these weak lines, 
there is a \ion{Ti}{i} line at 6707.964 \AA\ with $\chi$ = 1.88 eV and $\log gf=-6.903$, which 
appears in the VALD-2 database. This fact  probably led Reddy et 
al.~(\cite{red02}) to assume that the feature at 6708.025 \AA\ also belongs to Ti 
and has a similar excitation energy but much larger oscillator strength. For some reason 
Reddy et al.~(\cite{red02}) did not incorporate the Ti line at 6707.964 \AA\ in their 
final list. The two \ion{Ti}{i} lines at 6708.025 and 6708.125 \AA\  of Reddy et 
al.~(\cite{red02}) have never been tested in stars other than the Sun and 16 Cyg B. 

It is clear that the Li line list in the solar spectrum is not well
calibrated. Different authors obtain different adjusted $gf$ values
even when they use the same tools and spectra. Our goal is
not to make a better fit to the solar Li region since this does  
not guarantee a unique identification of the spectral lines. Instead,
we have decided to use the same list as Reddy et al.~(\cite{red02}) and see 
if we can explain observations of stars other than the Sun and 16 Cyg B. 

\begin{figure*}
\psfig{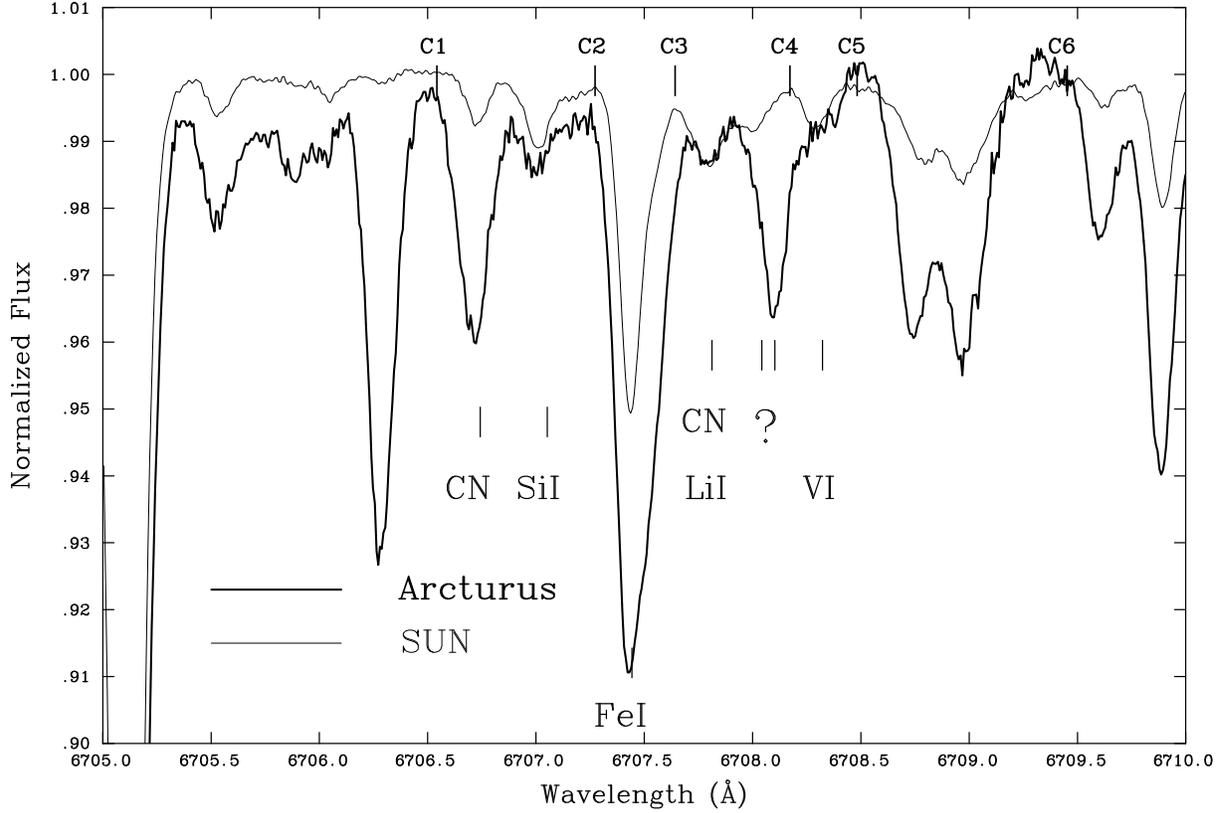}
\caption[]{High resolution and high S/N spectra of the Sun and Arcturus
in the Li region. The unidentified feature is centred on
6708.025 \AA\ in the Sun but appears at 6708.094 \AA\ in the spectrum 
of Arcturus.}
\label{fig1}
\end{figure*}

\begin{table*}
\caption[]{Abundance of \ion{Ti}{} (column 4) as derived from the EWs of two
 \ion{Ti}{ii} 
lines (columns 2 and 3) measured in our spectra. The EW of the unidentified feature 
(column 5) at 6708.025 \AA\ was computed assuming that it is due to the Ti line of 
Reddy et al.~(\cite{red02}) with abundances from column 4. Observed EWs of 
unidentified 
line are listed in column 5 with 3 $\sigma$ errors estimated using a Cayrel 
formulae 
(Cayrel \cite{cay88}). Values in parentheses in column 6 indicate that the line 
is not 
centred on 6708.025$\pm$0.01 \AA. The last two columns provide the Si abundance in our 
stars from 
Bodaghee et al.~(2003) and the EW of the unidentified feature assuming that it 
belongs 
to \ion{Si}{i} with $\chi$ = 6 eV and $\log gf=-2.97$. It is not clear if the
6708.025 \AA\ feature in HD\,200790 is real.}
\begin{tabular}{lccccccc}
\hline
\noalign{\smallskip}
Target  &  \ion{Ti}{ii}    & \ion{Ti}{ii}    & [Ti/H] & EW$_{\rm pred}$(\ion{Ti}{i}) 
& EW$_{\rm obs}$ & [Si/H] &
EW$_{\rm pred}$(\ion{Si}{i}) \\
        &  5418.77 \AA\ & 5336.78 \AA\ &          & 6708.025 \AA\ & 6708.025 \AA\   
	&  & 6708.025 \AA\  \\
\hline \\
HD\,1461   & 61.0 & 84.1  &   0.18  & 1.1    & 1.6$\pm$0.8 & 0.20  & 1.2 \\
HD\,13445  & 38.2 & 61.2  & $-$0.18  & 1.7    & (1.5$\pm$1.0) & $-$0.14 & 0.4 \\
HD\,16141  & 61.3 & 86.6  &   0.09  & 0.8   & 0.9$\pm$0.5 & 0.10  & 1.0 \\
HD\,22049  & 39.7 & 62.8  & $-$0.17  & 2.0    & (2.0$\pm$0.5) & $-$0.10 & 0.5 \\
HD\,75732A & 58.2 & 80.2  & 0.28    & 3.7    & (3$\pm$1.0)   & 0.38    & 1.5 \\
HD\,82943  & 61.5 & 82.4  & 0.27    & 0.9    &  -     & 0.27 & 1.3  \\
HD\,130322 & 45.7 & 69.8  & $-$0.06  & 1.4    & $<$ 1.5 & $-$0.04 & 0.7 \\
HD\,134987 & 62.6 & 85.2  & 0.3     & 1.4    & $<$ 1.5 & 0.27   &  1.3 \\
HD\,137759 & 86.8 & 109.6 & 0.03    & 14.0   & ($<$ 1.5) & 0.18 & 1.3  \\
HD\,145675 & 59.4 & 79.6  & 0.42    & 4.5    & (5.5$\pm$1) & 0.42 & 1.6 \\
HD\,177830 & 75.1 & 97.4  & 0.28    & 11.0   & ($<$ 2.0) & 0.45 & 1.9 \\
HD\,192263 & 43.2 & 64.6  & $-$0.01  & 4.0    & ($<$1.5)  & 0.02 & 0.5 \\
HD\,200790 & 63.5 & 89.7  & 0.02    & 0.3    & $<$0.6$>$  & 0.04 & 0.7 \\
HD\,210277 & 58.3 & 81.3  & 0.2     & 1.8    & 1.5$\pm$1.0 & 0.23 & 1.3 \\
HD\,217107 & 59.5 & 83.6  & 0.29    & 1.8    & 1.5$\pm$0.5 & 0.37 & 1.7 \\
\noalign{\smallskip}
\hline
\end{tabular}
\label{tab2}
\end{table*}

\section{The Li region in the Sun and in Arcturus}

The unidentified line at 6708.025 \AA\ is clearly seen in the solar spectrum (Fig. 1).
In the same figure we show a small spectral window from McDonald's 
atlas of a K III giant Arcturus observed with a resolving power 
150\,000 and S/N $\sim$ 1000 (Hinkle et al.~\cite{Hin00}). Most of the absorptions
in this small region appear in the spectra of both stars, which have very different
temperature, gravity and metallicity. It is interesting that the feature
at 6708.025 \AA\ does not become stronger in Arcturus. It may still 
exist in the spectrum but is severely blended with a strong absorption 
centred at 6708.094 \AA.  

The line at 6708.094 \AA\  is certainly not a redshifted \ion{Ti}{i} 6708.025 \AA\  
of Reddy et al.~(\cite{red02}). It also cannot be one of the TiO lines observed in 
much
cooler stars (Luck \cite{luck77}) 
and taking into account in our synthesis of three cool giants. 
The wavelength of this line corresponds to the \ion{V}{i} at 6708.094 \AA\  in the 
VALD-2 database 
and also appears in the list of Reddy et al.~(\cite{red02}) but with an oscillator 
strength that does not fit observations of Arcturus and other cool giants. 
Both wings of this line are disturbed by blends. The figure also 
demonstrates that blending is not the only problem in these studies. The exact 
location of the stellar continuum becomes a critical issue when we deal with 
absorption 
features as small as 1 m\AA.  

\begin{figure}
\psfig{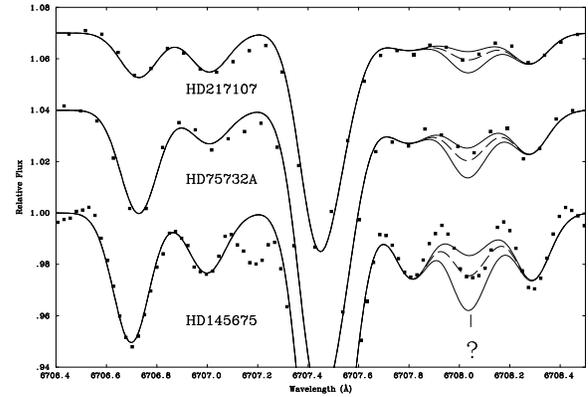}
\caption[]{Observed (filled small squares) and synthetic spectra of three 
metal-rich stars: 
HD\,217107, HD\,75732A and HD\,145675. The Ti abundances were computed for
[Ti/H] = 0 (thin continuous), 0.2 (dashed), 0.4 (thick continuous) in HD\,217107, 
[Ti/H] = 0 (thin continuous), 0.15 (dashed), 0.3 (thick continuous) in HD\,75732A and
[Ti/H] = 0 (thin continuous), 0.2 (dashed), 0.4 (thick continuous) in HD\,145675. 
}
\label{fig2}
\end{figure}

\begin{figure}
\psfig{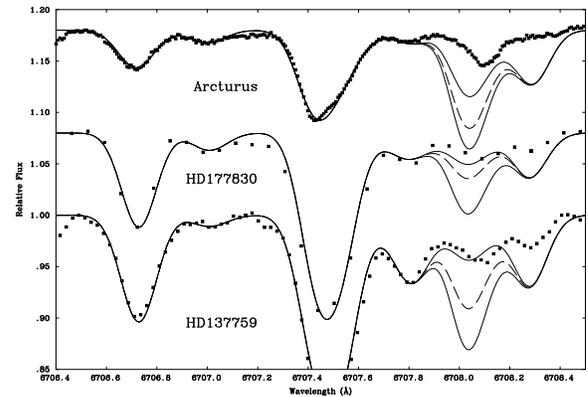}
\caption[]{Observed (filled small squares) and synthetic spectra of the 
Arcturus, HD\,177830 and HD\,137759. The Ti abundances were computed for
[Ti/H]  = $-$0.2 (thin continuous), $-$0.3 (dashed), $-$0.5 (thick continuous) in
 Arcturus, 
[Ti/H] = $-$0.2 (thin continuous), 0.0 (dashed), 0.3 (thick continuous) in HD\,177830 and
[Ti/H] = $-$0.4 (thin continuous), 0.0 (dashed), 0.4 (thick continuous) in HD\,137759.
}
\label{fig3}
\end{figure}

We know that echelle spectra are not calibrated in the absolute flux scale,
and that, therefore, the placement of local continuum levels is either visually 
estimated after fitting the extracted spectra with polynomials, or 
results from a detailed spectral synthesis. 
We have marked several points (C1-C6) in Fig.~1. which indicate very narrow windows 
theoretically free from spectral lines. These windows are used in metal-poor 
stars to make a fine tuning of the continuum and estimate the S/N. Points
C1, C2, C5 and C6 mark the solar continuum fitted by Kurucz et al.~(1984). 
We do not know how the continuum was placed in Arcturus but it appears from
the figure that only C5 and C6 can be considered as continuum points. 
Neither is it clear whether C2 and C4 are continuum points in the solar spectrum.  
The  point C3 lies on the continuum in those stars that do not have CN and Li 
absorptions at 6707.78 \AA. The strength of the unknown line at 6708.025 \AA\
and the \ion{V}{i} absorption at 6708.28 \AA\ define whether or not C4 is placed on  
the local continuum. The region between C2 and the \ion{Si}{i} line at 6707.01 \AA\  
is depressed in the solar spectrum as well as in the spectra of many other stars 
probably owing to some weak, still unknown, blends that start to appear in cool and
very metal-rich stars (e.g.~HD\,145675 in Fig.~2). Depending on the exact location 
of the points C1--C6 one may derive different EW values for the unidentified feature 
at 6708.025 \AA. However, as we will show below, even this uncertainty does 
not smear the huge differences between observations and spectrum synthesis 
when the Ti line of Reddy et al.~(\cite{red02}) is considered.

\section{The spectral feature at 6708.025 \AA}

\subsection{The \ion{Ti}{i} line of Reddy et al.~(2002)}

Let us investigate how the \ion{Ti}{i} 6708.025 \AA\ line of Reddy et 
al.~(\cite{red02}) 
fits our observations. We have computed Ti abundance in the sample of metal-rich 
stars (Table 1) using two lines of the 
ionized Ti from Santos et al.~(2000). The EWs of these lines and derived 
abundances are listed in Table 2. The difference in abundance from the
two lines was always less than 0.03 dex. Our tests with the neutral \ion{Ti}{i} 
lines from Santos et al.~(2000) showed that the abundances derived 
from \ion{Ti}{i} and \ion{Ti}{ii} agree within 0.1 dex in F8--G2 stars while there 
is a systematic difference at lower temperatures.  The abundances obtained 
from the \ion{Ti}{ii} lines are about 0.1--0.15 dex smaller than those 
derived using \ion{Ti}{i}. Because of a small non-LTE effect on \ion{Ti}{i}, 
the use of \ion{Ti}{ii} lines is preferred (N.~Shchukina 2002, personal 
communication). In addition, the abundances derived from \ion{Ti}{ii} lines 
are insensitive to the errors in $T_\mathrm{eff}$ (Santos et al.~2000). 
In any case, even if we suppose that 
our Ti abundances are uncertain by up to 0.3 dex (which is very unlikely, given
the errors considered in this work), this will not change our final 
conclusion regarding the reliability of the line proposed by Reddy et 
al.~(\cite{red02}). 

Using the abundances from the \ion{Ti}{ii} lines, we have estimated 
the EW of the feature at 6708.025 \AA, assuming that it belongs to the line 
proposed by Reddy et al.~(\cite{red02}). Predicted and observed EWs of the Ti line 
are listed in Table 2. We see an obvious disagreement at low $T_\mathrm{eff}$: 
the predicted line is much stronger than the observed one. In Fig.~2 we show spectrum 
synthesis for three metal-rich stars with different parameters. The abundances of
V, C, Si and other elements were scaled with Fe and then modified by small 
amounts (always less than 0.1 dex) in order to fit the observations. 
HD\,217107, with  $T_\mathrm{eff}=5655$ K, is more similar to the Sun  
than the two other stars and apparently for this reason the agreement between 
predicted and observed profiles of 6708.025 \AA\ is good. There is small room 
for making adjustments in continuum level in order to accommodate [Ti/H] = 0.3 and 
fit the observations. This plot alone cannot be used to argue against the 
identification proposed by Reddy et al.~(\cite{red02}). As a matter of fact, 
the observed and predicted EWs of the Ti line agree within errors in HD\,217107 
(Table 2). The discrepancy is larger in cooler objects HD\,145675 and HD\,75732A. 
We also note that the centre of the Ti line is slightly shifted to
the blue in the spectrum of HD\,145675. This shift and the disagreement 
between predicted and observed EW become much larger when we study 
K-type stars (Fig.~3). One of our targets is the well-known red giant 
Arcturus. The atmospheric parameters and the abundances of elements 
in this star were taken from the literature (Peterson et al.~1993). 
This figure demonstrates that the predicted Ti line becomes strong 
and its wavelength clearly does not correspond to the absorption 
observed at 6708.094 \AA. Although our computations rule out the [Ti/H] 
ratios $-$0.2 and $-$0.4 (thin continuous lines in Fig.~3) considered by us as
the lower limits in the synthesis of HD\,177830 and HD\,137759, we provided 
them in order to show the wavelength shift between the fictitious Ti 
line and the absorption at 6708.094 \AA. These plots clearly demonstrate
that the feature at 6708.025 \AA\ disappears at low temperatures
while a different line at 6708.094 \AA\  (not clearly observed in the solar
spectrum) becomes stronger. The wavelength of this line corresponds to
\ion{V}{i} in the VALD-2 database. Two lines at 6708.025 and 6708.094 \AA\  
appear as a single absorption feature in stars with $T_\mathrm{eff}$ = 5000--5400 K. 
It should be possible to separate them in very high resolution and high 
S/N spectra of stars with intrinsically narrow absorption lines.

When estimating 
the EW of the Ti line in HD\,82943 we found another disagreement with 
Reddy et al.~(\cite{red02}), who state that the combined strength of the 
two \ion{Ti}{i} lines (6708.025 and 6708.125 \AA) in the spectrum of 
HD\,82943 is 2.2 m\AA, while our analysis using exactly the same tools
and parameters shows 1.1 m\AA, in which the 6708.125 \AA\ line contributes
only 0.2 m\AA. The same disagreement appears when we consider the solar spectrum.
Using the list of Reddy et al.~(\cite{red02}),
 we found that the EWs of the \ion{Ti}{i} 
lines at 6708.025 and 6708.125 \AA\ are 0.75 and 0.17 m\AA, respectively. 
Thus, the combined EW = 0.92 m\AA\ is at odds with the 1.5 m\AA\  claimed by 
Reddy et al. (\cite{red02}).

\begin{figure}
\psfig{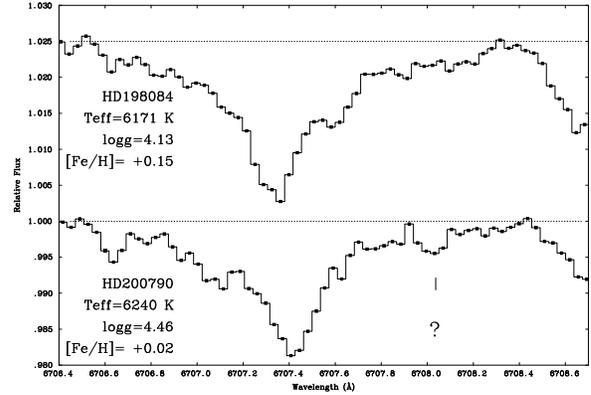}
\caption[]{The unidentified feature at 6708.025 \AA\ has EW $<$ 1 m\AA\ 
in HD\,198084 and HD\,200790.}
\label{fig4}
\end{figure}

\begin{figure}
\psfig{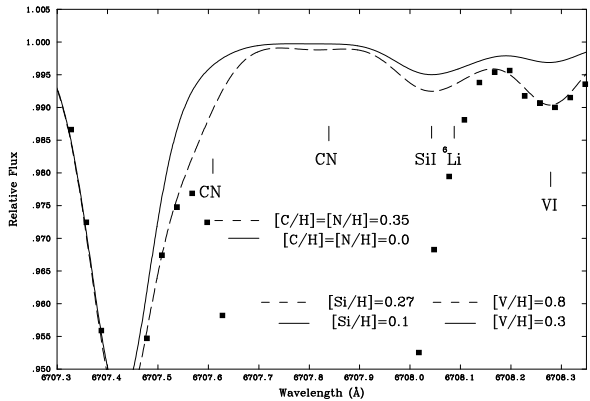}
\caption[]{The spectrum of HD\,82943 (small filled squares) and two synthetic spectra
computed without Li. The first synthesis (solid line) is computed without CN, 
and assuming [Si/H] = 0.1 and [V/H] = 0.3. Computations with [C/H] = [N/H] = 0.35,
[Si/H] = 0.27 and [V/H] = 0.8 are presented with a dashed line. The position of the
$^6$Li line is indicated.}
\label{fig5}
\end{figure}

\subsection{Identifying the line}

We have attempted a new identification of the 6708.025 \AA\ feature knowing 
that the strength of the line does not increase towards cooler temperatures. 
It is clear that any molecular or a low excitation line from a neutral 
metal will behave like the Ti line of Reddy et al.~(\cite{red02}). Since this is 
not observed, we are left with two possibilities. Either the transition 
belongs to an ion, or it is produced by a neutral element with a high excitation 
energy of the lower level.

The most abundant ions at these temperatures belong to Fe, Si, Mg, Ca and Ti. Reliable 
experimental line lists exist for all of these ions and  are available
in VALD-2. Most of the   \ion{Si}{ii}  linesin the Li region have very high excitation 
energies of between 10--16 eV. The same is true for \ion{Mg}{ii} 
($\chi$ = 11--13 eV) and \ion{Ca}{ii} ($\chi$ = 7--9.7 eV). \ion{Ti}{ii} has 
several transitions with $\chi$ = 1.9--3.1 eV. A \ion{Ti}{ii} line with 
$\chi$ = 1.9 eV 
and $\log gf=-4.1$ will reproduce the solar feature with EW = 0.75 m\AA. 
This line will also fit the spectra of the coolest objects in our sample, such 
as HD\,137759 and HD\,177830, where it will be weaker than 2 m\AA. In HD\,82943
it will have EW = 1.1 m\AA, which is not really different from the strength of
the neutral Ti line (0.9 m\AA). However, small discrepancies may appear in hotter
stars. For example, given the Ti abundance in HD\,200790, we anticipate about 
1 m\AA\   absorption which we think is not observed in the high S/N spectrum 
(Fig.~4). 
The marginal feature observed at 6708.028 \AA\  is about 0.6 m\AA\ and it is hard to 
judge by eye if it is real. On the upper panel of the same graph we show the 
spectrum  of HD\,198084. Unfortunately, this star is in a binary
system and we cannot make any strong statements regarding the strength of
the unidentified absorption. However, we would still expect to see the blend 
of two unidentified lines at 6708.025 \AA. In fact, there is a broad absorption 
around 6708.0 \AA,  most probably caused by  the shifted unidentified lines but its
EW does not exceed 1 m\AA.

We have studied a spectrum of the hot Li-poor star 
HR 7697 ($T_\mathrm{\rm eff}$=6588 K, $\log g$ = 4,24, [Fe/H] = 0.01) with  
S/N $\sim$ 550 kindly provided by Ann Boesgaard (Boesgaard et al.~2001).
The ionized line of Ti in this object would produce about 1 m\AA\ absorption, which
we could not find in the data. These tests make the idea of an ionized Ti line
less attractive, although, given the errors in the data and our analysis, we
still cannot completely rule out this possibility. We also note that it will be 
much harder to identify the line if it belongs to an ion of an element 
whose abundance does not scale with Fe.

Rare Earth elements were also considered. The best candidates are perhaps 
two \ion{Ce}{ii} lines listed in the DREAM database 
(http://www.umh.ac.be/~astro/dream.shtml) 
with wavelengths 6708.077 and 6708.099 \AA. The lines are weak ($\log gf=-2.57$ and 
$\log gf=-2.12$) and have excitation energies $\chi_{l}$ = 2.25 and 0.7 eV,
 respectively. 
Let us suppose that the 6708.077 \AA\ line is responsible for the feature at 
6708.025 \AA. 
The  6708.077 \AA\  absorption is absolutely negligible in the solar spectrum, and in 
order to produce 0.7 m\AA\  it must have $\log gf=0.43$. This very large correction 
is not 
allowed by accurate radiative lifetime measurements (Zhang et al.~2001). If we 
still suppose that the adjustment is possible, these two lines will produce about 
4 m\AA\ absorption in Arcturus. However, the synthetic spectra will not reproduce 
the observed profile at 6708.094 \AA\  unless we modify the wavelengths. In brief,
we can hardly force these two lines of \ion{Ce}{ii} to explain observations. 

\begin{table*}
\caption[]{Mean abundance [Si/H] = 0.27 $\pm$ 0.02 was computed from 11 Si lines 
observed in S/N$\sim$1000 VLT/UVES and WHT/UES spectra 
of HD\,82943.}
\begin{tabular}{lcccc}
\hline
\noalign{\smallskip}
Wavelength (\AA)   &  $\chi$   &    $\log gf$    &   EW(m\AA)   &  [Si/H] \\
\hline \\
5665.56 &  4.92 &   $-$1.980	 &   53.5   & 0.27    \\
5690.43 &  4.93 &   $-$1.790	 &   62.8   & 0.22   \\
5701.10 &  4.93 &   $-$2.020	 &   52.2   & 0.30   \\
5772.14 &  5.08 &   $-$1.620	 &   69.8   & 0.28   \\
5793.09 &  4.93 &   $-$1.910	 &   57.3   & 0.26   \\
5948.55 &  5.08 &   $-$1.110     &  109.7   & 0.26   \\
6125.02 &  5.61 &   $-$1.520	 &   46.1   & 0.26   \\
6142.49 &  5.62 &   $-$1.480	 &   50.9   & 0.29   \\
6145.02 &  5.61 &   $-$1.400	 &   54.9   & 0.25   \\
6155.15 &  5.62 &   $-$0.750     &  113.0   & 0.27   \\
6721.86 &  5.86 &   $-$1.090     &   65.4   & 0.26   \\

\noalign{\smallskip}
\hline
\end{tabular}
\label{tab3}
\end{table*}

Our synthetic spectra also demonstrate that the $gf$ values of 
\ion{V}{i} lines at 6708.094 and 6708.280 \AA\ in 
the list of Reddy et al.~(\cite{red02}) do not explain observations of
HD\,137759, HD\,177830 and Arcturus (Fig.~3) assuming that the 
abundance of V scales with Fe (Sadakane et al.~2002). Our tests show that even 
adopting 
[V/Fe] = $-$0.5 in these stars, we still cannot fit the feature at 6708.280 \AA. 
Such low abundances of V are excluded from the analysis of other spectral lines.
On the other hand, the fit to the same line in HD\,82943 was achieved by 
boosting the abundance to [V/H] = 0.8. This high value is ruled out from 
a synthesis of different \ion{V}{i} lines. We faced similar problem for the line at 
6708.094 \AA\ but in the opposite direction. The newly added \ion{Ti}{i} line at 
6708.125 \AA\ in the list of Reddy et al.~(\cite{red02}) cannot salvage the situation 
as it appears too weak in our targets. We also note that the \ion{V}{i} line 
at 6708.280 \AA\  does not appear in the VALD-2 database (Kupka et al.~1999).

\subsection{The \ion{Si}{i} line of M\"{u}ller et al.~(1975)}

Given the large number of high excitation Si lines in the Li region
(see VALD-2 database), we have decided to test if the Si line at 6708.025 \AA\ 
proposed by 
M\"{u}ller et al.~(\cite{mull75}) can explain our observations. In fact, 
the high excitation Si line that appears at 6707.01 \AA\  (Fig.~1) is very 
similar to the one proposed by M\"{u}ller et al.~(\cite{mull75}). It behaves exactly 
in the same way in stars with different temperatures as the absorption at 
6708.025 \AA.
Given the EW (= 0.75 m\AA)  of the 6708.025 \AA\ feature in the solar spectrum, 
we have estimated $\log gf=-2.97$  for the \ion{Si}{i} line with 
$\chi_{l}$ = 6 eV proposed
by M\"{u}ller et al.~(\cite{mull75}). Our $gf$ value is slightly different from that
of M\"{u}ller et al.~(\cite{mull75}) owing to the differences in the model atmospheres 
and the adopted abundance of Si (Anders \& Grevesse \cite{And89}),
 which is 0.05 larger 
than the value used by M\"{u}ller et al.~(\cite{mull75}). We have computed the Si 
abundance in our stars using the lines from  Bodaghee et al.~(2003) listed in Table 3. 
The new list is an 
extended and improved version of the one considered by Santos et al.~(\cite{San00}). 
The abundance errors were always less than 0.07 dex and  do not affect our 
analysis. Predicted EWs of the Si line are listed in the last column. As we can 
see, there is good agreement between observations and synthesis. 

In conclusion, we think it is more likely that the spectral feature
at 6708.025 \AA\ belongs to a high excitation transition of some neutral
atom. The small spectral window near 6708 \AA\  contains many high excitation Si
lines; therefore, the identification proposed by M\"{u}ller et al.~(\cite{mull75}) 
seem to be the best choice. However, we cannot rule out the possibility 
that the 6708.025 \AA\ line belongs to an ion or to a rare 
element. More high quality observations are necessary to tackle this problem.

\section{The $^6$Li/$^7$Li  ratio in HD\,82943}

HD\,82943 was newly observed at VLT/UVES with a resolving power 105\,000 and a
signal-to-noise ratio close to 1000. Reanalysis of the $^6$Li/$^7$Li ratio 
has been carried out using the same tools as in Israelian et al.~(2001) and the 
line list of Reddy et al.~(\cite{red02}). The only change we have made in this 
list is a replacement of the \ion{Ti}{i} line at 6708.025 \AA\ by the high excitation 
\ion{Si}{i} line with $\log gf=-2.97$  and $\chi_l$ = 6 eV. This change does not 
help in giving a higher $^6$Li/$^7$Li ratio in HD\,82943 since the EW of the
\ion{Si}{i} line in HD\,82943 is even larger than that of the Ti
of Reddy et al.~\cite{red02} (see Table 2). Furthermore, an accurate 
abundance of Si in HD\,82943 needs to be calculated. For this purpose we have 
used eleven Si lines from Table 3 and computed their EWs using
very high S/N spectra obtained with VLT/UVES and WHT/UES. In Table 3
we list the lines, measurements and resulting abundances. We found 
[Si/H] = 0.27 $\pm$ 0.02 in excellent agreement with the value given by 
Sadakane et al.~(\cite{sad02}) after accounting for differences in stellar 
parameters.

The identification of the chemical element is
also important, although Reddy et al.~(\cite{red02}) have ignored this point by 
stating 
that the identification of the pair as \ion{Ti}{i} is not critical to the 
analysis. The EWs of the Ti line of Reddy et al.~(\cite{red02}) are 0.45 and 
0.9 m\AA\ in HD\,82943 for [Ti/H] = 0 and [Ti/H] = 0.3, respectively. This factor
 of 2 
difference in the EWs is equal to 0.05 correction in $^6$Li/$^7$Li ratio
if the total EW of the Li line is 10 m\AA. This is a significant difference,
and given the fact that in many stars [Ti/H] can be larger than 0.3,
the abundance correction may affect the $^6$Li/$^7$Li ratios by at least 
0.03 dex, depending on the strength of the $^7$Li line.

\begin{figure}
\psfig{width=\hsize,file=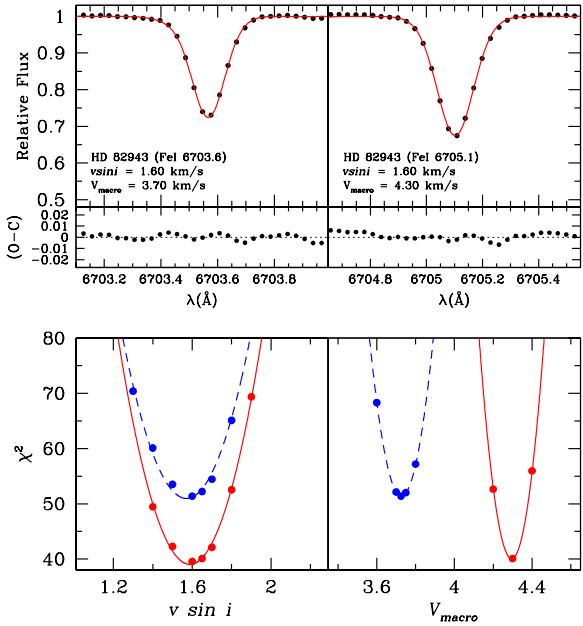}
\caption[]{Observed and computed profiles of two Fe lines and results of $\chi^2$ 
analysis. Filled dots in the upper panel are observations and continuous lines are
synthetic spectra derived from the $\chi^2$ analysis. Dashed and solid lines in 
the lower panel are the $\chi^2$ analysis of \ion{Fe}{i} 6703.6 and 6705.1 \AA\ 
lines, respectively.}
\label{fig6}
\end{figure}

\begin{figure}
\psfig{width=\hsize,file=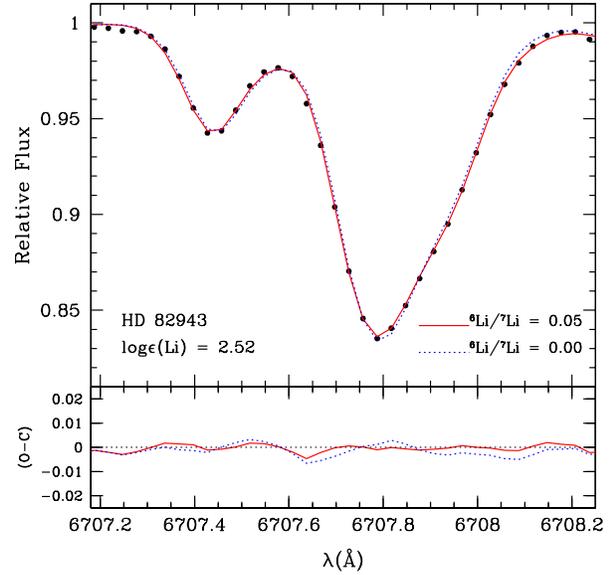}
\caption[]{Comparison of the observed (filled large dots) and synthetic spectra of 
HD\,82943 corresponding to $^6$Li/$^7$Li = 0.05 (continuous) and $^6$Li/$^7$Li = 0 
(dotted) 
isotopic ratios. They correspond to the best fits with $f$($^6$Li) = 0 and a
wavelength offset of $-$0.65\,km\,s$^{-1}$ (small dots) and to the f($^6$Li) = 0.05
 with a
wavelength offset of $-$0.44\,km\,s$^{-1}$ (continuous line). Fits to the blue wing 
of
the Li profile can be improved if we adopt the wavelengths of CN lines from 
Brault \& M\"{u}ller (1975). The CN lines observed in the arc spectrum by 
these authors appear at 6707.55 \AA\ while Reddy et al.~(2002) and
Lambert et al.~(1993) list them between 6707.464 and 6707.529 \AA.
The residuals (O-C) of the observations after subtraction of the 
synthetic spectra are shown.}
\label{fig7}
\end{figure}

To demonstrate the blending of the Si and CN lines in the Li feature 
of HD\,82943, we have computed synthetic spectra without considering
any of the Li isotopes. Figure 5 shows that  CN and Si make a significant
contribution to the total EW of the Li line. Given the high C and N 
content of HD\,82943 (Santos et al.~\cite{San00}, Sadakane et al.~\cite{sad02})
one may consider the abundances of these elements as free parameters in the 
$\chi^2$ analysis. 
The strength of the Si line may affect the synthesis as well. It is 
also important to realize that this line cannot exactly mimic the $^6$Li 
component at 6708.0728 \AA\  because their wavelengths differ by 0.05 \AA. 

In obtaining the broadening parameters we used the same \ion{Fe}{i} lines
as in our first paper (Israelian et al.~\cite{isr01}). From the $\chi^2$ analysis of 
two Fe lines (Fig.~6) we obtained a mean $v\sin i$ = 1.6 $\pm$ 0.05\,km\,s$^{-1}$  and 
$V_{\rm mac}$ = 4 $\pm$ 0.3\,km\,s$^{-1}$, in good agreement with our previous results
and also with those of Reddy et al.~(\cite{red02}). The synthetic spectra were 
also convolved with a 
Gaussian function representing the instrumental profile. Given the low
rotational velocity in HD\,82943, it is possible to obtain fits using simple
Gaussian functions. 
As we have noticed before (Israelian et al.~\cite{isr01}, see also 
Reddy et al.~\cite{red02}), the final values of $^6$Li/$^7$Li 
are not affected by using a pure Gaussian broadening function. Average 
broadening parameters obtained from the Fe lines were used to derive 
the $^6$Li/$^7$Li ratio.

\begin{figure}
\psfig{width=\hsize,file=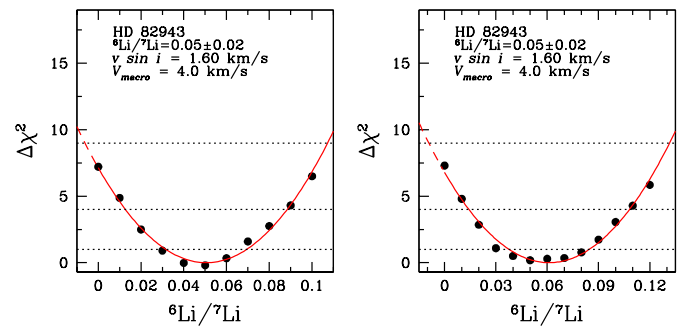}
\caption[]{Results from the $\chi^2$ analysis for the $^6$Li/$^7$Li ratio. 
$\Delta \chi^2$=1,4,and 9 correspond to 1,2 and 3$\sigma$ confidence 
limit is (dotted lines). Continuum adjustments were allowed 
within 0.2 $\%$ (left panel) and 0.5$\%$ (right panel).}
\label{fig8}
\end{figure}

The free parameters in a standard  $\chi^2$ analysis of the $^6$Li/$^7$Li ratio 
in HD\,82943 were the total abundance of Li, the wavelength shift, 
the $^6$Li fraction and the position of the continuum. 
The final $^6$Li/$^7$Li ratio is not strongly  affected by the blue CN absorption,
and after few a iterations we fixed the CN abundance equal to [C/H]  = [N/H] = 0.35. 
Small adjustments of the continuum level not exceeding 0.2$\%$ (given the S/N 
of the spectrum) were allowed to optimize the fits. The abundance of the Si line was 
also considered as a free parameter in the range [Si/H] = 0.27 $\pm$ 0.1 but after a few 
iterations it was fixed to [Si/H] = 0.27. Additional computations
were carried out allowing for large variations of $v\sin i$ and $V_{\rm mac}$
exceeding the 3$\sigma$ limit derived from the Fe lines. We found that these
effects do not change our final results. The minimum $\chi^2$ was derived by 
optimizing all free parameters for each value. The new analysis using 
Reddy et al.'s (\cite{red02}) line list and the Si line yields $f$($^6$Li) = 0.05 $\pm$ 0.02 
(Fig.~8, left panel). 
The best fit for $f$($^6$Li) = 0 has an rms of 2.869 $\times$ 10$^{-3}$, which is significantly
larger than the rms (9.91 $\times$ 10$^{-4}$) of the best fit with $f$($^6$Li) = 0.05.
Wavelength offsets in the best fits for $f$($^6$Li) = 0 and $f$($^6$Li) = 0.05 
were $-$0.65\,km\,s$^{-1}$ and $-$0.44\,km\,s$^{-1}$, respectively. These values
 agree perfectly with those reported by Cayrel et al.~(1999) in their $^6$Li/$^7$Li analysis 
of the halo star HD\,84937. The synthetic and the observed profiles for two cases 
are shown in Fig.~7. Given the strength of the Li line (about 50 m\AA), these shifts are
consistent with observations of Fe lines in the solar spectrum 
(Allende Prieto \& Garc\'\i a L\'opez \cite{all98}). Since the scatter of the velocity 
shifts of Fe lines in the solar spectrum is large (ranging from 
$-$0.8 to 0.2\,km\,s$^{-1}$), one cannot make any predictions for the blueshifts 
of the Li lines based on a statistical trend observed for the solar Fe lines. 
So the best option still remains to keep the wavelength shift of the Li line 
as a free parameter. 

The minimum $\chi^2$ is not affected by large variations in the continuum either.
Relaxing the upper limit of continuum change from 0.2 to 0.5 $\%$ does not change 
our conclusion (Fig.~8, right panel). The final value of $f$($^6$Li) is also not 
sensitive to the Si abundance. If we repeat the analysis fixing the continuum but
changing the abundance of Si, we will find that the best fit to the observed 
profile is achieved for [Si/H] = 0.27 (Fig.~9, right panel). The same analysis for  
$f$($^6$Li) = 0 yields [Si/H] = 0.51 (Fig.~9, left panel). Such a high abundance of Si 
is ruled out by our data. However, even if we assume that it is possible, 
the quality of the fit when $f$($^6$Li) = 0.05 and [Si/H] = 0.27 is about 4 times better 
(i.e.~$\chi^2$ is smaller) compared with the case when $f$($^6$Li) = 0.0 and 
[Si/H] = 0.51. On the other hand, the \ion{Si}{i} line may suffer from convective 
blueshifts as well. To account for this effect, we have repeated the $\chi^2$ 
analysis considering the wavelength of the Si line as another free parameter. 
We found that the effect is very small. For example, when $f$($^6$Li) = 0.05, the 
$-$8 m\AA\ wavelength shift will introduce a negligible change in $\chi^2$ of about 
5$\%$. This would also imply a small increase in the Si abundance equal to [Si/H] = 0.31. 
These changes do not affect our final results.

\begin{figure}
\psfig{width=\hsize,file=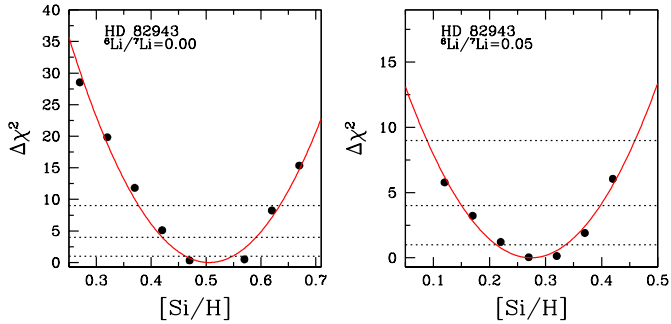}
\caption[]{Results from the $\chi^2$ analysis for the abundance of Si.}
\label{fig9}
\end{figure}

Reddy et al.~(\cite{red02}) noted in 
their paper that the Ti lines contribute 2.2 m\AA\  (while
our own estimate is EW = 1.1 m\AA) into the 50 m\AA\ Li line, which, according to
these authors, corresponds to $^6$Li/$^7$Li = 0.06. Given that
without this blend the ratio was $^6$Li/$^7$Li = 0.12 (Israelian et al.~\cite{isr01}), 
one should wonder what is causing the remaining 0.06 contribution.
If the contribution of the blend is only 0.06, then we are still
left with another $^6$Li/$^7$Li = 0.06. The blended line is not 
strong enough  to completely eliminate the contribution from $^6$Li. 
Although Reddy et al.~(\cite{red02}) have stated that our spectra agree 
very well, they did not explore this puzzle.

 In order to judge the validity of our spectral synthesis and the
placement of the stellar continuum, we have synthesized spectra of $\alpha$ Cen A and 
HD\,82943 in the 7 \AA\ region near the Li doublet (Fig.~10). Observations of 
$\alpha$ Cen A  with a resolving power of 105.000 are taken from Rebolo et al.~(1986) 
and have S/N $>$500 in the Li region. As stellar parameters for $\alpha$ Cen A we
have adopted  $T_\mathrm{eff}$ = 5800 K, $\log{g}$ = 4.31, $\xi_{t}$ = 1.1 km\,s$^{-1}$, 
[Fe/H] = 0.24 (Chmielewski et al.~1992) and abundances were taken from the literature 
(Abia et al.~1988; Santos et al.~2000; Chmielewski et al.~1992; King et al.~1997).

\begin{figure}
\psfig{width=\hsize,file=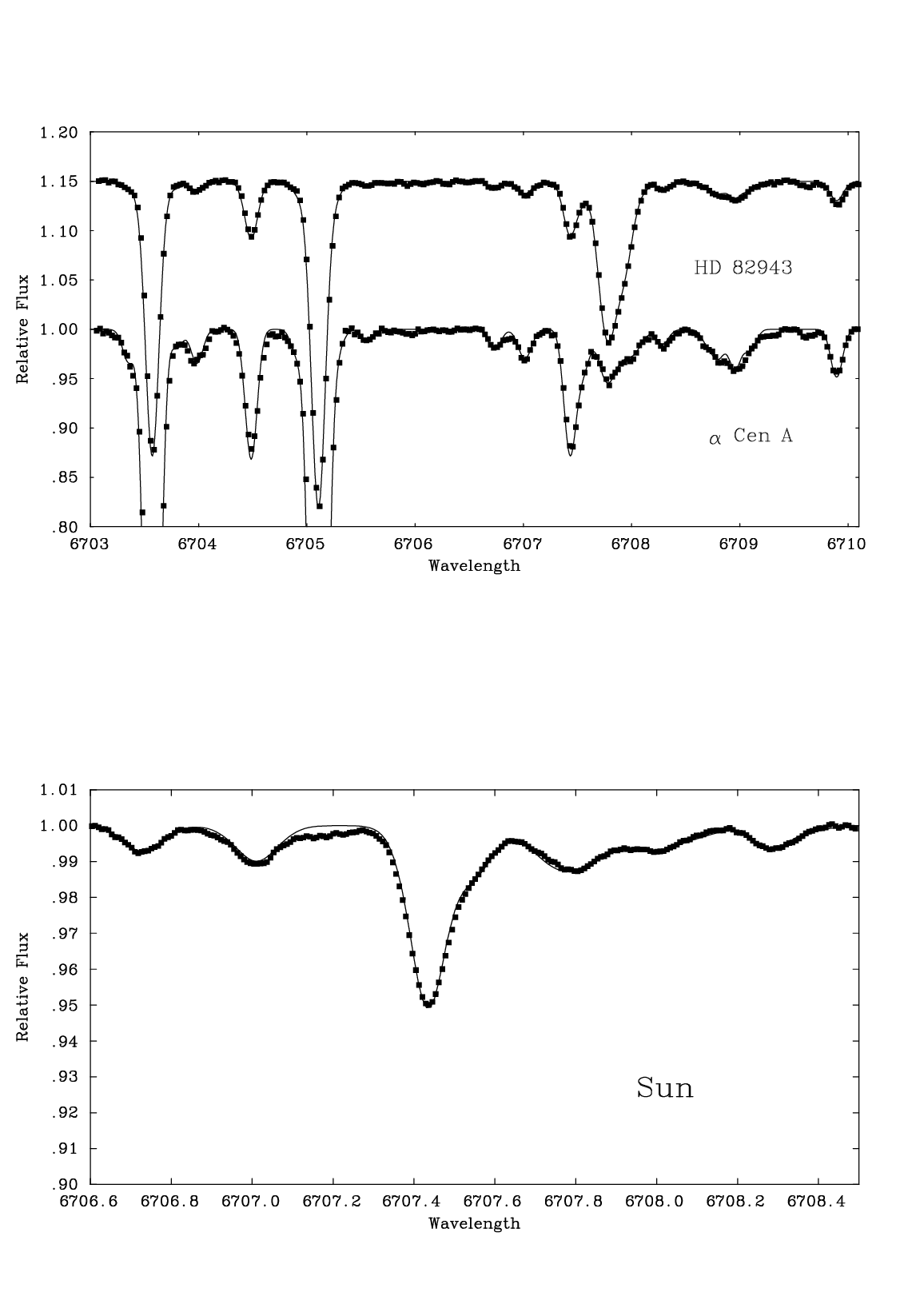,angle=0}
\caption[]{Upper panel: comparison of the observed (filled squares) and synthetic spectra of 
HD\,82943 and $\alpha$ Cen A. Lower panel: comparison of the observed (filled squares, 
Kurucz et al.~1984) and synthetic spectra of the Sun using the list of Reddy et al.~with
the wavelengths of CN lines from Brault \& M\"{u}ller (1975) and the Si line discussed in
this article. There are missing lines at 6707.2 with a total EW $<$ 0.3 m\AA\ 
(see also Reddy et al.~2002; Nissen et al.~1999; M\"{u}ller et al.~1975), which are clearly 
seen for the first time in the spectrum of a very metal rich cool star HD\,145675 (Fig.~2). 
These weak lines do not affect our analysis (i.e.~the continuum placement) as they disappear 
in stars with $T_\mathrm{eff}$ $>$ 5900 K.}
\label{fig10}
\end{figure}

Given  that we are not able to make a unique identification
of the chemical element responsible for the blend, the $^6$Li/$^7$Li 
found from the analysis cannot not be considered to be a final value. 
The abundance of Si usually scales with Fe in planet host stars 
(see for example Santos et at.~\cite{San00}; Gonzalez et al.~\cite{Gon01}; 
Sadakane et al.~\cite{sad02}; Takeda et al.~\cite{tak01}). However, if the spectral 
line belongs to an element which is under- or overabundant with respect 
to Fe, then the $^6$Li/$^7$Li ratio will be different. In most cases 
the elemental abundances are not very different from that of Fe (typically within $\pm$0.1 dex) 
and therefore the overall effect on the $^6$Li/$^7$Li ratio will not be large.

\section{Conclusions}

High quality observations of metal-rich stars with undetectable Li have
essential consequences, first for the identification of weak lines in
the Li 6708 \AA\ region, and second for the analysis of the $^6$Li/$^7$Li isotopic 
ratio.
Our results clearly rule out the \ion{Ti}{i} line at 6708.025 \AA\ suggested by 
Reddy et al.~(\cite{red02}). Alternatively, we propose that this feature can be
attributed to the \ion{Si}{i} line of M\"{u}ller et al.~(1975).  
The identification and the presence of another \ion{Ti}{i} line at 6708.125 
\AA\ in Reddy et al. (2002) line list is also not justified and cannot be 
verified with our dataset as this line is
extremely weak in all our targets. We also find that two lines of \ion{V}{i}
at 6708.094 and 6708.280 \AA\  from the same list have incorrect parameters.  
However, these \ion{V}{i} lines and the \ion{Ti}{i} line at 6708.125 \AA\ cannot 
strongly affect the $^6$Li/$^7$Li ratio in stars with $T_\mathrm{eff} > $ 5500 K.

Our analysis of the new  VLT/UVES spectra (S/N=1000), taking into consideration the 
\ion{Si}{i} line at 6708.025, gives an isotopic ratio $^6$Li/$^7$Li = 0.05$\pm$0.02 
in HD\,82943. This implies a contribution to the total absorption 
in the Li region  of 1.3 m\AA\ and 2.7 m\AA\ due to \ion{Si}{i} and $^6$Li, respectively.  
If no \ion{Si}{i} blend  is considered in the analysis
we obtain $^6$Li/$^7$Li=0.08, which is $\sim$ 33 $\%$ smaller than the value 
obtained in our first analysis (Israelian et al.~2001) based on a spectrum with S/N=500. 
Since $^6$Li can be construed as a fossil record of rocky matter accretion during
planetary system formation, further high quality observations of this fragile 
isotope may offer us a unique and invaluable insight into the past dynamical history 
of extra-solar planetary systems.

\begin{acknowledgements}

We would like to thank Drs Roger Cayrel, Yoichi Takeda, Poul Eric Nissen, Johannes Andersen, 
Eric Sandquist, Lew Hobbs,  Eswar Reddy and David Lambert for 
several helpful discussions. Ann Boesgaard
has kindly provided us with her spectra of Li-poor stars. We also thank 
Jonay Gonzalez Hernandez for reducing several WHT/UES spectra used in this article
and Carlos Allende Prieto for providing us with the McDonald Atlas of Arcturus.  
We are grateful to the referee Dr Ruth Peterson for several helpful comments and
suggestions.

\end{acknowledgements}

\end{document}